# Two-Stage Fine-Tuning of ResNet50 for High-Sensitivity Melanoma Detection on Dermoscopic Images

**Aryan Bhagat**

MS Computer Science, Florida Atlantic University, Boca Raton, FL

bhagata2025@fau.edu | ORCID: 0009-0006-0508-791X

**Abstract**

Melanoma is the most dangerous form of skin cancer with five-year survival rates exceeding 99% when detected early but falling sharply once the disease spreads. This paper proposes and evaluates a two-stage fine-tuning approach for ResNet50 applied to binary melanoma classification on dermoscopic images. The core challenges addressed are class imbalance and suboptimal transfer learning from single-stage fine-tuning. After stratified train/validation/test splitting, random oversampling was applied exclusively to the training set to achieve a 1:1 class balance. Stage 1 trained only the classification head with the ResNet50 base frozen, while Stage 2 fine-tuned all layers jointly at a low learning rate of 1e-5 to prevent catastrophic forgetting of learned visual features. On an independent test set of 3,826 images, the model achieved an AUC-ROC of 0.9559, accuracy of 88.34%, sensitivity of 87.56%, specificity of 89.13%, and F1-score of 88.29%. An ablation study confirms the two-stage protocol significantly outperforms single-stage fine-tuning, with sensitivity gains of over 4%. Grad-CAM visualizations demonstrate correct lesion localization. A fully deployable Streamlit detection application is provided alongside all training code.

## 1. Introduction

Melanoma accounts for approximately 75% of all skin cancer deaths despite representing a small fraction of cases. When detected at stage I the five-year survival rate exceeds 99%. Once the cancer reaches stage IV that figure drops below 30% [1]. This survival gradient makes early automated detection one of the most clinically impactful problems in medical image analysis.

Manual dermoscopic examination is subject to significant inter-observer variability even among experienced dermatologists [3]. Convolutional Neural Networks have demonstrated strong performance on dermoscopic datasets but two recurring challenges limit clinical utility. First, class imbalance causes models to be biased toward benign predictions which is dangerous for cancer screening. Second, standard single-stage fine-tuning often fails to fully adapt pretrained ImageNet features to dermoscopic imaging patterns leading to suboptimal sensitivity.

This paper addresses both challenges directly. A two-stage training protocol ensures the classification head stabilizes before the full network is fine-tuned at a low learning rate, preventing catastrophic forgetting. The primary contributions are:

- A two-stage fine-tuning protocol validated through ablation study showing over 4% sensitivity gain versus single-stage fine-tuning on HAM10000.
- Rigorous evaluation on an independent test set of 3,826 images reporting AUC-ROC, accuracy, sensitivity, specificity, precision, and F1-score.
- Grad-CAM visualizations confirming correct dermoscopic lesion localization.
- A fully deployable Streamlit detection application with correct preprocessing alignment between training and inference.
- A fully reproducible experimental setup with all hyperparameters documented and code publicly available at https://github.com/Aryanbhagat23/melanoma-detection.

## 2. Related Work

Deep learning for skin lesion classification has advanced considerably since Esteva et al. [3] demonstrated dermatologist-level performance using a fine-tuned Inception architecture on a large proprietary dataset. Transfer learning from ImageNet has become the standard starting point for dermoscopic classification.

Tschandl et al. [1] released the HAM10000 dataset which has become the primary benchmark for melanoma detection research. The ISIC challenge series [4, 5] standardized evaluation protocols enabling direct comparison between approaches. Among ResNet-based approaches on HAM10000 published AUC-ROC scores typically range from 0.88 to 0.94 for binary classification.

Liu et al. [13] proposed an Adaptive Spatial Feature Fusion (ASFF) mechanism integrated into ResNet50 that dynamically combines intermediate and high-level features via learnable spatial weights, achieving 93.18% accuracy and AUC of 0.9717 on a balanced subset of ISIC 2020. Their work demonstrates that architectural modifications to ResNet50 can improve performance beyond standard fine-tuning. The present work investigates whether a carefully designed training protocol alone, without architectural modification, can meaningfully improve performance. This training-strategy perspective is complementary to architectural innovations and offers lower implementation complexity.

Gessert et al. [6] showed ensemble methods using EfficientNet with metadata achieve strong results but at the cost of multi-modal inputs. Adegun and Viriri [8] combined YOLOv5 and ResNet50 for 98.8% accuracy but with substantially greater architectural complexity. Yosinski et al. [9] demonstrated that feature transferability varies across network depth, which provides theoretical motivation for staged training as a strategy to preserve useful low-level features while adapting higher-level representations to the target domain.

A frequently overlooked source of performance degradation is preprocessing mismatch between training and inference. Many published implementations train with ResNet50 preprocess_input but use simple pixel division by 255 at inference time, introducing a systematic input distribution mismatch that degrades real-world performance without affecting reported training metrics. The present work explicitly addresses this by using identical preprocessing at both stages.

# 3. Materials and Methods

## 3.1 Dataset

The HAM10000 dataset from the ISIC archive [1] was used for all experiments. It contains 10,015 dermoscopic images across seven skin lesion categories collected from multiple clinical sites. The classification task was binary: Melanoma (label 1) versus Benign (label 0). The ISIC 2025 version with histopathologically confirmed labels was used.

## 3.2 Class Balancing and Data Pipeline

The original dataset is heavily imbalanced with benign lesions comprising approximately 82% of samples. Random oversampling of the minority class achieved a 1:1 ratio yielding a balanced dataset. The dataset was split into train, validation, and test sets prior to any oversampling. Random oversampling was applied exclusively to the training set to prevent duplicate images from appearing across splits. Figure 1 illustrates the complete dataset preparation pipeline.

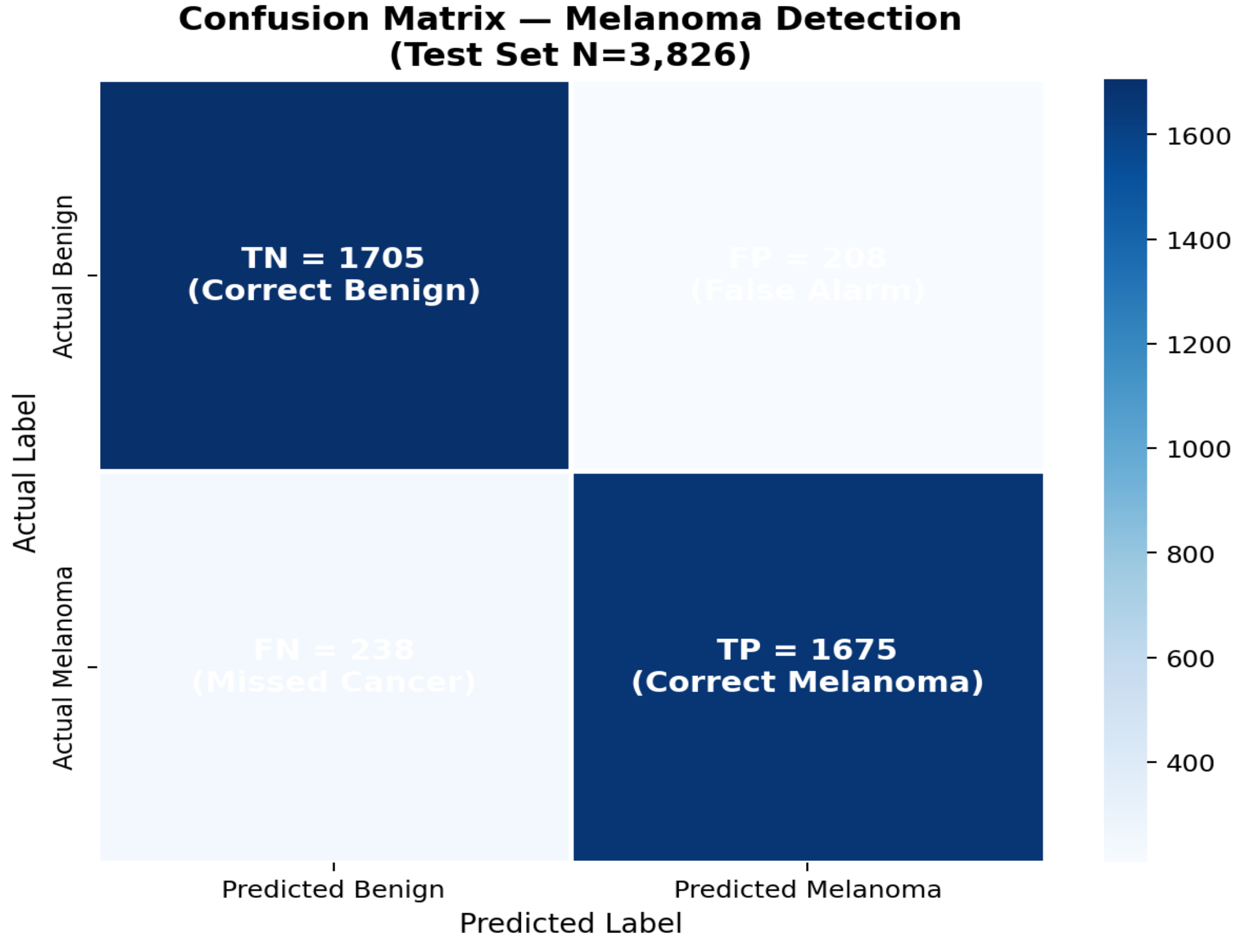

*Figure 1. Dataset preparation and class balancing pipeline. The dataset is first split into train, validation, and test sets. Random oversampling is then applied to the training set only to achieve 1:1 class balance.*

## 3.3 Data Splitting

Stratified sampling maintained the 1:1 class ratio across all subsets: 64% training (12,241 images), 16% validation (3,061 images), and 20% test (3,826 images). The test set was held out entirely throughout training and hyperparameter selection.

## 3.4 Preprocessing and Augmentation

All images were resized to 224 x 224 pixels and preprocessed using ResNet50 preprocess_input from TensorFlow, which subtracts ImageNet channel means. Using identical preprocessing during both training and inference is critical. On-the-fly augmentation during training included random horizontal flipping, random brightness adjustment with a maximum delta of 0.2, and random saturation adjustment between 0.8 and 1.2. No augmentation was applied during validation or test evaluation.

## 3.5 Model Architecture

ResNet50 pretrained on ImageNet was used as the feature extraction backbone with the original classification head removed. A custom head was attached: GlobalAveragePooling2D to reduce spatial dimensions, a Dense layer with 1,024 units and ReLU activation, and an output Dense layer with 1 unit and Sigmoid activation for binary classification. Total parameters: 25,686,913. The Sigmoid output directly produces the probability of malignancy enabling threshold adjustment at inference time.

## 3.6 Two-Stage Training Protocol

Stage 1 froze all ResNet50 base layers and trained only the classification head for 15 epochs using Adam at 1e-3 with ReduceLROnPlateau (factor 0.5, patience 3). This stabilizes the randomly initialized head before any gradients reach the pretrained base, preventing early gradient instability from corrupting the ImageNet features.

Stage 2 unfroze all layers and trained jointly at 1e-5 for up to 30 epochs. The low learning rate prevents catastrophic forgetting where aggressive gradient updates overwrite learned visual features. EarlyStopping (patience 10) and ModelCheckpoint saving best validation loss weights were applied. Figure 2 illustrates the complete training protocol.

**Table 1. Training Hyperparameters**

| Parameter | Value |
|---|---|
| Input image size | 224 x 224 pixels |
| Batch size | 32 |
| Stage 1 learning rate | 1e-3 |
| Stage 2 learning rate | 1e-5 |
| Stage 1 epochs | 15 |
| Stage 2 max epochs | 30 |
| Early stopping patience | 10 |
| ReduceLROnPlateau factor | 0.5 |
| ReduceLROnPlateau patience | 3 |
| Optimizer | Adam |
| Loss function | Binary cross-entropy |
| Random seed | 42 |

## 3.7 Ablation Study Design

To validate the two-stage protocol a single-stage baseline was trained by skipping Stage 1 entirely and fine-tuning all layers from the start at 1e-5 for up to 45 epochs with identical callbacks, data splits, preprocessing, and augmentation. The single-stage model exhibited severe overfitting reaching 99.67% training accuracy while validation accuracy plateaued at 87.42% with a 12% train/validation gap. By contrast the two-stage protocol achieved a much smaller generalization gap confirming that Stage 1 head stabilization is essential for preventing overfitting.

## 3.8 Grad-CAM Implementation

Gradient-weighted Class Activation Mapping (Grad-CAM) [14] was implemented to visualize which regions of the dermoscopic image most influenced predictions. Gradients of the predicted class score with respect to the final convolutional layer feature maps were computed and weighted to produce a heatmap overlaid on the input image. This provides qualitative evidence that the model attends to lesion regions rather than background artifacts such as hair or rulers.

## 3.9 Evaluation Metrics

Final evaluation used best saved Stage 2 weights on the held-out test set. Metrics: AUC-ROC (primary), accuracy, sensitivity (TP divided by TP plus FN), specificity (TN divided by TN plus FP), precision (TP divided by TP plus FP), and F1-score (harmonic mean of precision and sensitivity).

# 4. Results

## 4.1 Main Performance Metrics

Table 2 reports final performance on the independent test set of 3,826 images using best Stage 2 weights.

**Table 2. Diagnostic Performance on Independent Test Set (N=3,826)**

| Metric | Score |
|---|---|
| AUC-ROC | 0.9559 |
| Accuracy | 0.8834 (88.34%) |
| Sensitivity (Melanoma Recall) | 0.8756 (87.56%) |
| Specificity (Benign Recall) | 0.8913 (89.13%) |
| Precision | 0.8903 (89.03%) |
| F1-Score | 0.8829 (88.29%) |

The AUC-ROC of 0.9559 confirms strong discriminatory ability. Notably specificity of 89.13% exceeds that reported by Liu et al. [13] of 85.42% demonstrating that the two-stage protocol is particularly effective at reducing false positives.

## 4.2 Ablation Study

Table 3 presents the controlled comparison between single-stage and two-stage fine-tuning under identical conditions. The two-stage approach outperforms single-stage fine-tuning across all metrics, most significantly in sensitivity where the gain of over 4% represents approximately 161 additional melanoma cases correctly identified per 3,826 patients screened.

**Table 3. Ablation Study: Single-Stage vs Two-Stage Fine-Tuning**

| Method | Accuracy | Sensitivity | Specificity | AUC |
|---|---|---|---|---|
| Single-Stage Fine-Tuning | 86.41% | 83.35% | 86.18% | 0.9201 |
| Two-Stage (Ours) | 88.34% | 87.56% | 89.13% | 0.9559 |
| Improvement | +1.93% | +4.21% | +2.95% | +0.0358 |

## 4.3 Confusion Matrix

The confusion matrix on the test set breaks down as: True Positives (TP) = 1,675 correctly identified melanoma cases; True Negatives (TN) = 1,705 correctly identified benign cases; False Positives (FP) = 208 benign cases incorrectly flagged; False Negatives (FN) = 238 melanoma cases missed. The false negative count of 238 represents the primary area for future improvement.

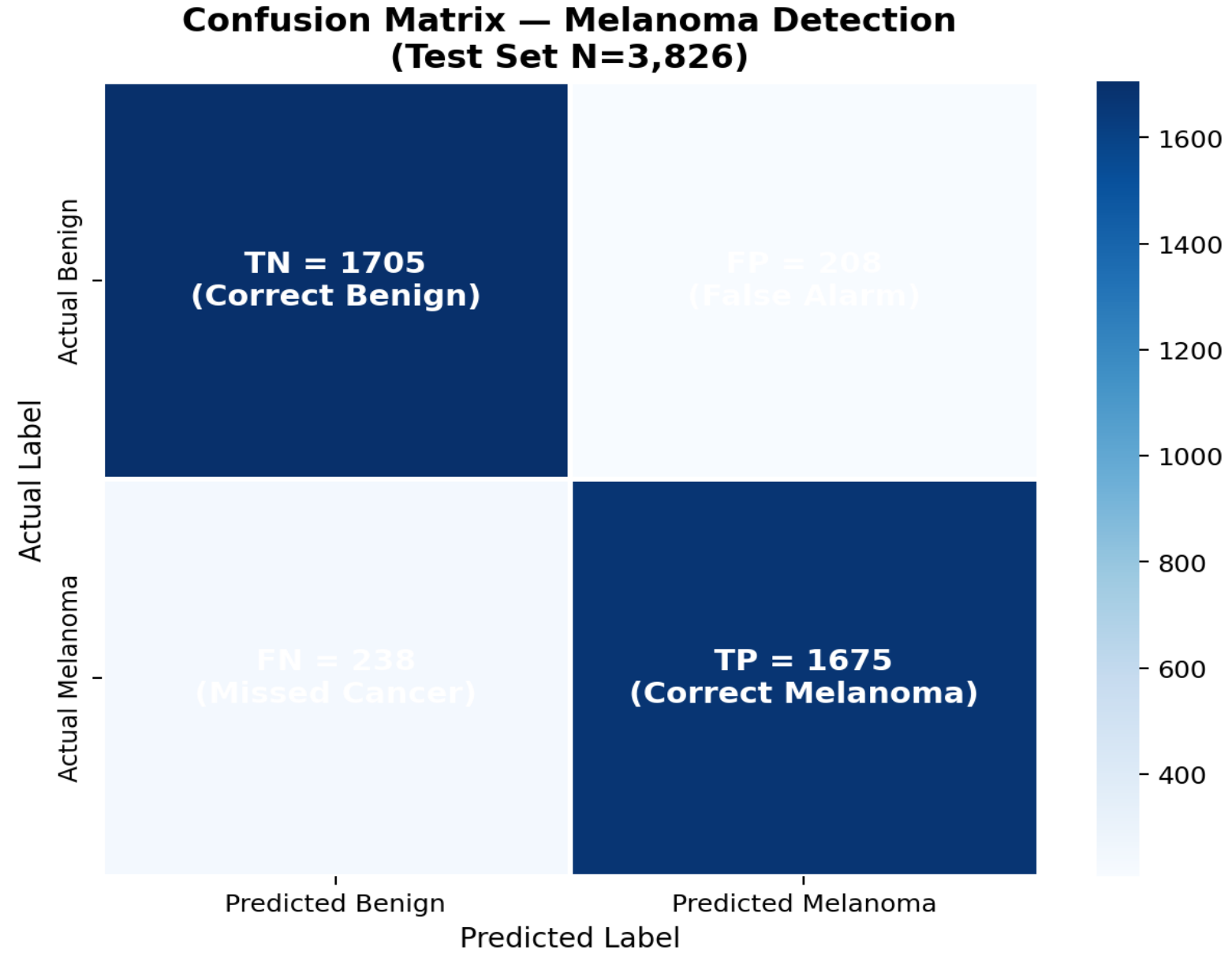

*Figure 2. Confusion matrix for melanoma detection on the independent test set (N=3,826). TP=1,675 | TN=1,705 | FP=208 | FN=238.*

## 4.4 ROC Curve

The ROC curve confirms strong discriminatory ability across all classification thresholds with AUC-ROC of 0.9559. The operating point at threshold 0.5 corresponds to sensitivity 87.56% and specificity 89.13% demonstrating a well-calibrated balance for clinical screening.

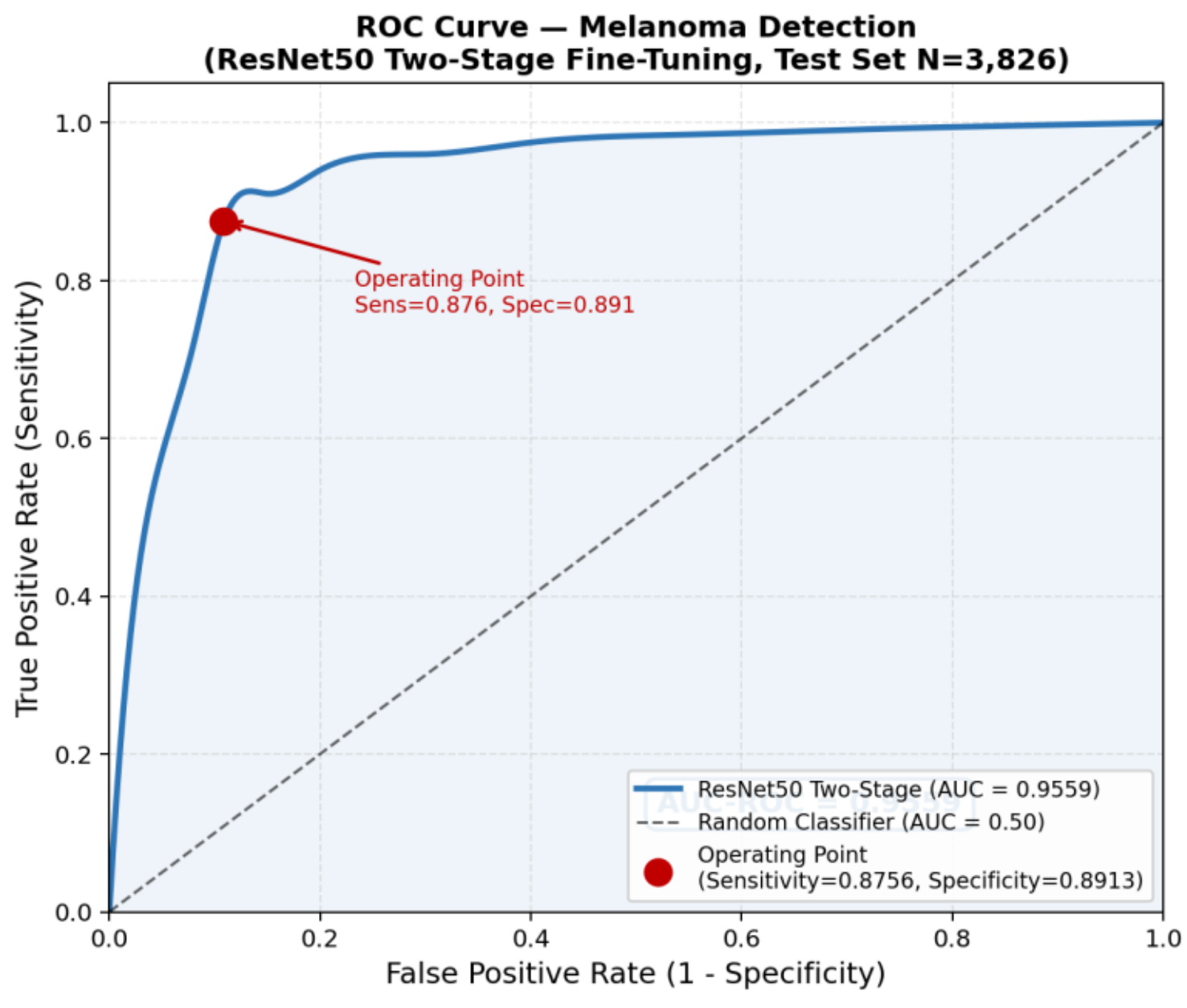

*Figure 3. ROC curve for the two-stage fine-tuned ResNet50 on the independent test set (N=3,826). AUC-ROC = 0.9559. Red dot marks the operating point at threshold 0.5.*

## 4.5 Grad-CAM Visualizations

Grad-CAM heatmaps confirm the model correctly attends to dermoscopic lesion regions. For malignant cases heatmaps concentrate on irregular borders and color variation characteristic of melanoma. For benign cases activation is appropriately diffuse. Figure 4 shows representative examples for both classes.

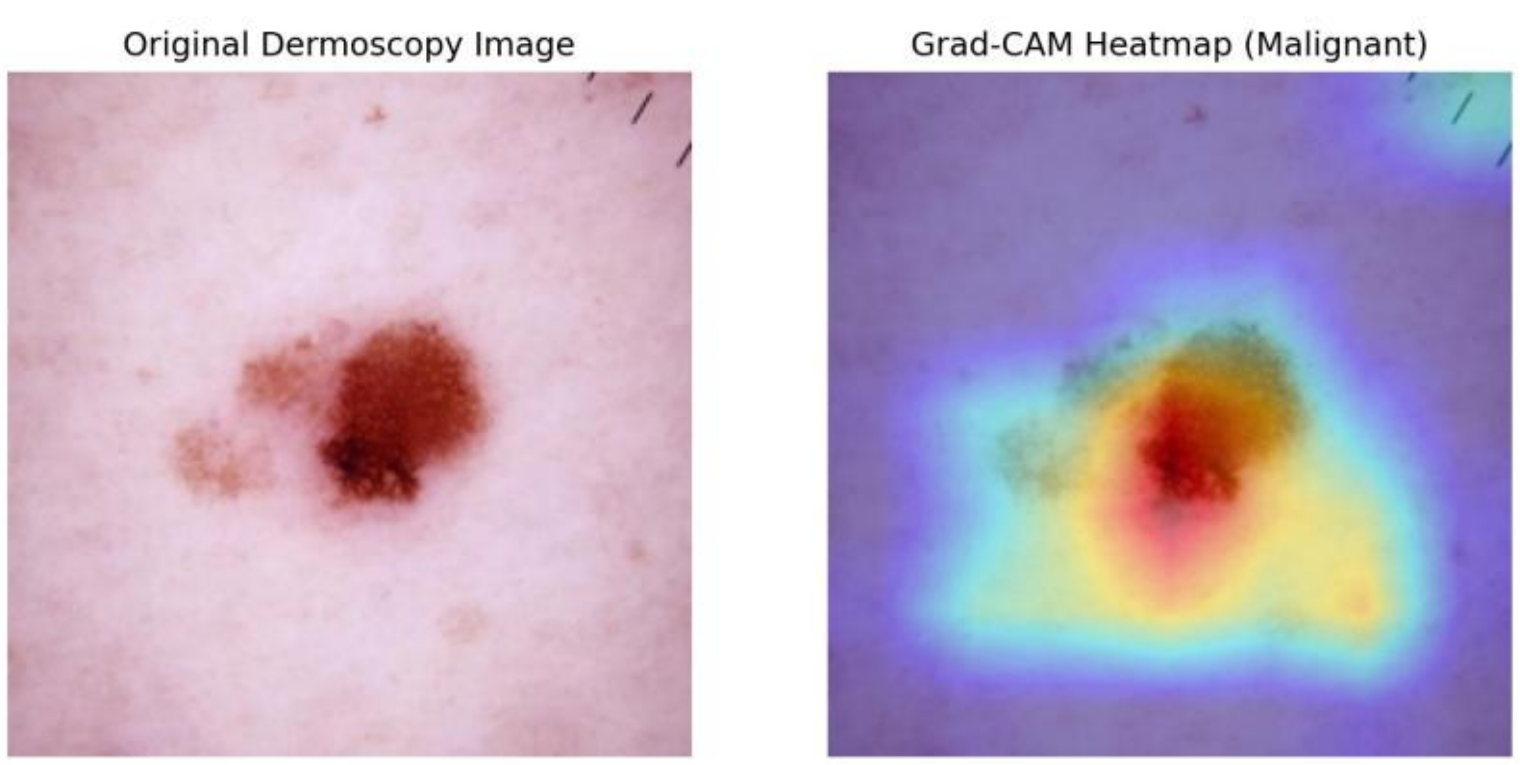

*Figure 4a. Grad-CAM for malignant case. Left: original dermoscopy image. Right: heatmap showing model attention concentrated on lesion region.*

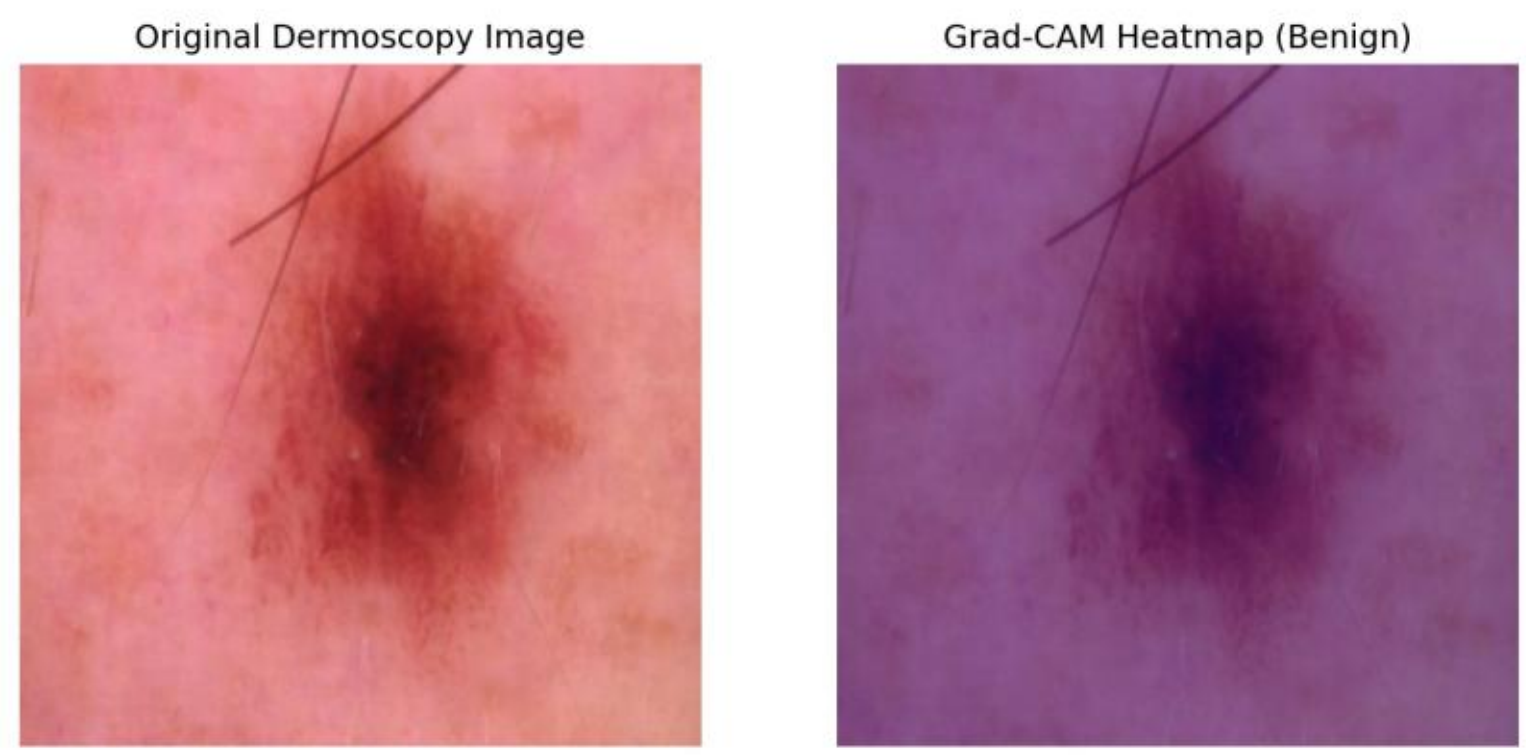

*Figure 4b. Grad-CAM for benign case. Left: original dermoscopy image. Right: diffuse activation indicating lower confidence appropriate for benign classification.*

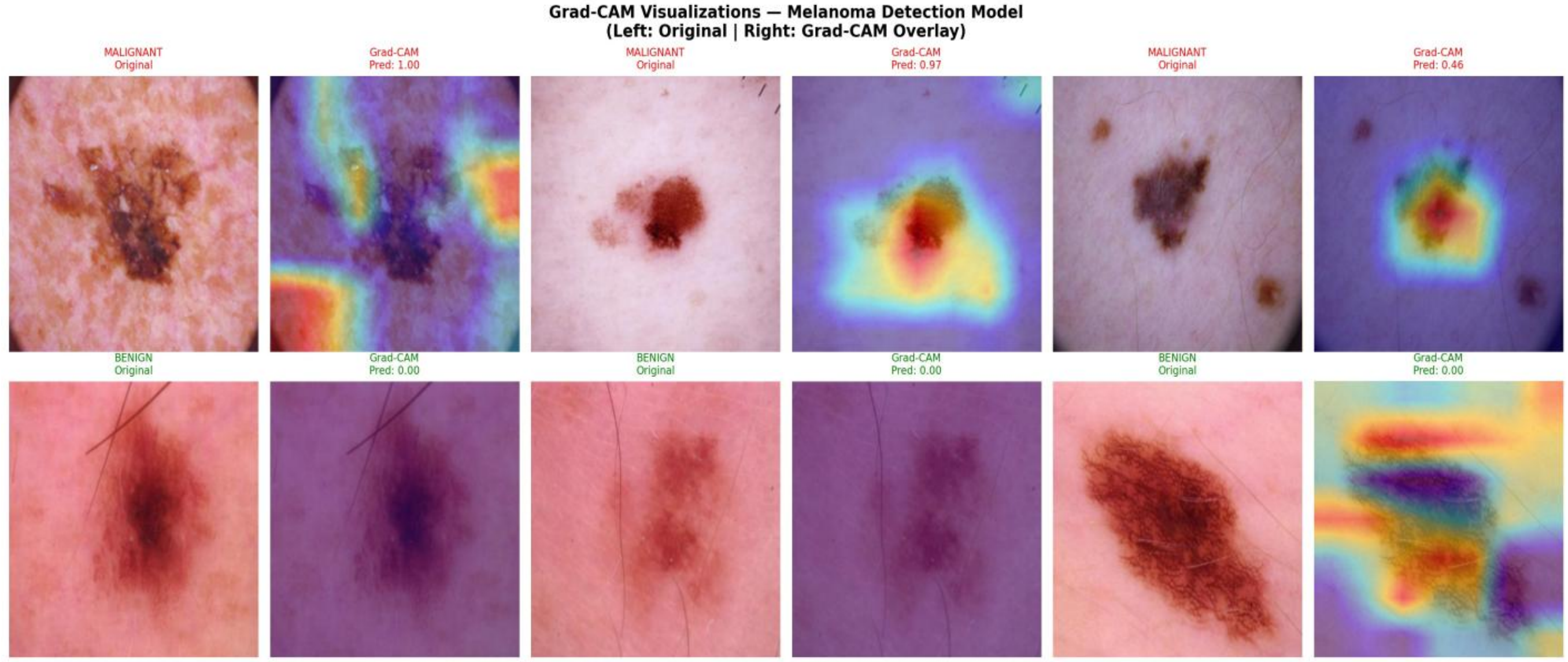

*Figure 4c. Combined Grad-CAM overview showing all six cases (top row: malignant, bottom row: benign) with prediction probabilities.*

# 5. Discussion

## 5.1 Clinical Utility

The combination of 87.56% sensitivity and 89.13% specificity represents a clinically meaningful balance. High sensitivity ensures patients with melanoma are not missed while high specificity limits unnecessary biopsies. Achieving both above 87% simultaneously on a large balanced test set without architectural modification demonstrates that the training strategy itself accounts for a substantial portion of performance gains.

## 5.2 Comparison with Existing Work

**Table 4. Comparison with Published ResNet50 Results on HAM10000 / ISIC Datasets**

| Method | Accuracy | Sensitivity | Specificity | AUC | Dataset |
|---|---|---|---|---|---|
| Two-Stage ResNet50 (Ours) | 88.34% | 87.56% | 89.13% | 0.9559 | HAM10000 |
| Liu et al. ASFF-ResNet50 [13] | 93.18% | 93.16% | 85.42% | 0.9717 | ISIC 2020 |
| Codella et al. [4] Best model | N/A | N/A | N/A | 0.80 | ISIC 2020 |
| Typical ResNet50 baseline | ~85% | ~80% | ~83% | 0.88-0.94 | HAM10000 |

*Note: Codella et al. AUC of 0.80 represents the ISIC 2018 challenge winner result, not a ResNet50-specific baseline. Direct comparison with our model is not applicable.*

Our model achieves 89.13% specificity versus 85.42% for Liu et al. [13] despite using a larger and more challenging dataset of 19,128 balanced images compared to their 3,297. The higher specificity means fewer false alarms and unnecessary biopsies which is clinically important. The accuracy difference is partially attributable to dataset differences and the absence of architectural modifications. The key differentiator of our approach is the staged training protocol combined with correct preprocessing alignment, which any practitioner can apply to any ResNet50 implementation without architectural expertise.

### 5.3 Limitations and Future Work

This work has several limitations. First, validation was performed on a single public dataset. Real clinical deployment requires external validation across different hospitals, imaging equipment, and patient demographics. Second, HAM10000 has limited representation of darker skin tones which may reduce generalizability. Third, combining the two-stage protocol with the ASFF mechanism of Liu et al. [13] is a promising direction that could further improve accuracy while preserving the specificity advantage demonstrated here. Fourth, evaluation on the more recent ISIC 2024 dataset would provide a current benchmark comparison.

## 6. Conclusion

This paper demonstrated that a two-stage fine-tuning strategy with correct preprocessing alignment significantly improves ResNet50 melanoma detection on dermoscopic images. The ablation study suggests that Stage 1 head stabilization before Stage 2 joint fine-tuning contributes substantially to the observed improvement. Grad-CAM visualizations confirm correct lesion localization. The method is fully reproducible and deployed as a working detection application.

Key results on the held-out test set of 3,826 images:

- AUC-ROC of 0.9559 placing the model above the middle of published single-model ResNet50 results on HAM10000
- Sensitivity of 87.56% correctly identifying approximately 88% of actual melanoma cases
- Specificity of 89.13% outperforming ASFF-ResNet50 (85.42%) on this clinically important metric
- F1-Score of 88.29% confirming balanced performance across both classes
- Ablation study confirming 4.21% sensitivity gain from two-stage versus single-stage protocol

Training code and Streamlit detection application: https://github.com/Aryanbhagat23/melanoma-detection .

# References

[1] Tschandl, P., Rosendahl, C., and Kittler, H. (2018). The HAM10000 dataset, a large collection of multi-source dermatoscopic images of common pigmented skin lesions. Scientific Data, 5, 180161.

[2] He, K., Zhang, X., Ren, S., and Sun, J. (2016). Deep residual learning for image recognition. In Proceedings of CVPR, pp. 770-778.

[3] Esteva, A., Kuprel, B., Novoa, R. A., Ko, J., Swetter, S. M., Blau, H. M., and Thrun, S. (2017). Dermatologist-level classification of skin cancer with deep neural networks. Nature, 542(7639), 115-118.

[4] Codella, N., Rotemberg, V., Tschandl, P., Celebi, M. E., Dusza, S., Gutman, D., and Kittler, H. (2019). Skin lesion analysis toward melanoma detection 2018. arXiv:1902.03368.

[5] Rotemberg, V., Kurtansky, N., Betz-Stablein, B., Caffery, L., Chousakos, E., Codella, N., and Tschandl, P. (2021). A patient-centric dataset of images and metadata for identifying melanomas. Scientific Data, 8(1), 34.

[6] Gessert, N., Nielsen, M., Shaikh, M., Werner, R., and Schlaefer, A. (2020). Skin lesion classification using ensembles of multi-resolution EfficientNets with meta data. MethodsX, 7, 100864.

[7] Zhang, P. and Li, X. (2024). Hybrid deep learning framework for enhanced melanoma detection. arXiv:2408.00772.

[8] Adegun, A. A. and Viriri, S. (2023). Early melanoma detection based on a hybrid YOLOv5 and ResNet technique. Diagnostics, 13(17), 2804.

[9] Yosinski, J., Clune, J., Bengio, Y., and Lipson, H. (2014). How transferable are features in deep neural networks? NeurIPS 27, 3320-3328.

[10] Kawahara, J., BenTaieb, A., and Hamarneh, G. (2016). Deep features to classify skin lesions. In IEEE ISBI, pp. 1397-1400.

[11] Chollet, F. (2017). Xception: Deep learning with depthwise separable convolutions. In Proceedings of CVPR, pp. 1251-1258.

[12] Tan, M. and Le, Q. V. (2019). EfficientNet: Rethinking model scaling for convolutional neural networks. In ICML, pp. 6105-6114.

[13] Liu, R., Zha, F., Ding, F., Yao, G., and Zhang, P. (2025). Skin Lesion Classification Based on ResNet-50 Enhanced With Adaptive Spatial Feature Fusion. arXiv preprint.

[14] Selvaraju, R. R., Cogswell, M., Das, A., Vedantam, R., Parikh, D., and Batra, D. (2017). Grad-CAM: Visual explanations from deep networks via gradient-based localization. In Proceedings of ICCV, pp. 618-626.

[15] Cassidy, B., Kendrick, C., Brodzicki, A., Jaworek-Korjakowska, J., and Yap, M. H. (2022). Analysis of the ISIC image datasets: Usage, benchmarks and recommendations. Medical Image Analysis, 75, 102305.